\documentclass[twoside]{article}
\usepackage[super,sort&compress,comma]{natbib} 
\usepackage[version=3]{mhchem}
\usepackage{times}
\usepackage{sectsty}
\usepackage{balance} 

\usepackage{graphicx} %eps figures can be used instead
\usepackage{lastpage}
\usepackage[format=plain,justification=raggedright,
singlelinecheck=false,font=small,labelfont=bf,labelsep=space]{caption} 
\usepackage{fancyhdr}
\usepackage{amsmath}
\usepackage{amssymb}
\usepackage{datetime}
\begin{document}

\thispagestyle{plain}

% \fancypagestyle{plain}{
% \fancyhead[L]{\includegraphics[height=8pt]{headers/LH}}
% \fancyhead[C]{\hspace{-1cm}\includegraphics[height=20pt]{headers/CH}}
% \fancyhead[R]{\includegraphics[height=10pt]{headers/RH}\vspace{-0.2cm}}
% \renewcommand{\headrulewidth}{1pt}}
 \renewcommand{\thefootnote}{\fnsymbol{footnote}}
 \renewcommand\footnoterule{\vspace*{1pt}% 
 \hrule width 3.4in height 0.4pt \vspace*{5pt}} 
 \setcounter{secnumdepth}{5}

\makeatletter 
\def\subsubsection{\@startsection{subsubsection}{3}{10pt}{-1.25ex plus -1ex 
minus -.1ex}{0ex plus 0ex}{\normalsize\bf}} 
\def\paragraph{\@startsection{paragraph}{4}{10pt}{-1.25ex plus -1ex minus 
-.1ex}{0ex plus 0ex}{\normalsize\textit}} 
\renewcommand\@biblabel[1]{#1}
\renewcommand\@makefntext[1]% 
{\noindent\makebox[0pt][r]{\@thefnmark\,}#1}
\makeatother 
\renewcommand{\figurename}{\small{Fig.}~}
\sectionfont{\large}
\subsectionfont{\normalsize} 

% \fancyfoot{}
% \fancyfoot[LO,RE]{\vspace{-7pt}\includegraphics[height=9pt]{headers/LF}}
% \fancyfoot[CO]{\vspace{-7.2pt}\hspace{12.2cm}\includegraphics{headers/RF}}
% \fancyfoot[CE]{\vspace{-7.5pt}\hspace{-13.5cm}\includegraphics{headers/RF}}
% \fancyfoot[RO]{\footnotesize{\sffamily{1--\pageref{LastPage}
% ~\textbar\hspace{2pt}\thepage}}}
% \fancyfoot[LE]{\footnotesize{\sffamily{\thepage~\textbar\hspace{3.45cm}
% 1--\pageref{LastPage}}}}
% \fancyhead{}
% \renewcommand{\headrulewidth}{1pt} 
% \renewcommand{\footrulewidth}{1pt}
% \setlength{\arrayrulewidth}{1pt}
% \setlength{\columnsep}{6.5mm}
% \setlength\bibsep{1pt}

\twocolumn[
\begin{@twocolumnfalse}
\noindent\LARGE{\textbf{Towards a dynamical approach to the calculation of the 
figure of merit of thermoelectric nanoscale devices}}
\vspace{0.6cm}

\noindent\large{\textbf{Roberto D'Agosta,\textit{$^{a,b,\ddag}$}}}\vspace{0.5cm}
%Please note that \ast indicates the corresponding author(s) but no footnote 
%text is required. 
%\\
%\currenttime~~\today 

%\noindent\textit{\small{\textbf{Received Xth XXXXXXXXXX 20XX, Accepted Xth 
%XXXXXXXXX 20XX\newline
%First published on the web Xth XXXXXXXXXX 200X}}}

%\noindent \textbf{\small{DOI: 10.1039/b000000x}}
\vspace{0.6cm}
%Please do not change this text.

\noindent \normalsize{Research on thermoelectrical energy conversion, the reuse
of waste heat produced by some mechanical or chemical processes to generate
electricity, has recently gained some momentum. The calculation of the
electronic parameters entering the figure of merit of this energy conversion,
and therefore the discovery of efficient materials, is usually performed
starting from the Landauer's approach to quantum transport coupled with the
Onsager's linear response theory. As it is well known, that approach suffers of
certain serious drawbacks. Here, we discuss alternative dynamical methods that
can go beyond the validity of the Landauer's/Onsager's approach for electronic
transport. They can be used to validate the predictions of the
Landauer's/Onsager's approach and to investigate systems for which that
approach has shown to be unsatisfactory.}

\vspace{0.5cm}
 \end{@twocolumnfalse}
]

%%%%%%%%%%%%%%%%%%%%%%%%%
\footnotetext{\textit{$^{a}$ETSF Scientific Development Center,
Departamento de F\'isica de Materiales, Universidad del Pa\'is Vasco, E-20018
San Sebasti\'an, Spain}}
\footnotetext{\textit{$^{b}$IKERBASQUE, Basque Foundation for Science, E-48011, 
Bilbao, Spain}}

%additional addresses can be cited as above using the lower-case letters, c, d, 
%e... If all authors are from the same address, no letter is required
\footnotetext{$\ddag$~Email: roberto.dagosta@ehu.es}

\section{Introduction} Electrical energy is one of the most versatile forms of
energy, since it can be easily stored, moved, or converted to any other form
(work, heat, etc.). On the other hand, in the process of electrical energy
production, a relevant part of the energy, indeed as much as 60$\%$ in most
cases, is wasted as heat, e.g., the plumes from the cooling towers of a power
station. The efficiency of this conversion process is limited by fundamental
laws and, nonetheless, it is difficult to do much better than the present
values. Nonetheless, no physical principle forbids us from using part of the
waste heat of the first energy conversion to obtain more energy. Indeed, if the
conversion process were ideally efficient no energy could be obtained through
any other physical or chemical process. In reality, for non ideal cycles, by
reusing the waste heat, we will pay a price in wasted energy, but looking at
the overall process we will end up producing more electrical energy and less
waste heat. This two-step process is limited in efficiency by thermodynamical
laws, but it is still more efficient than the single step process. Many
alternative methods to convert that waste heat back to energy have been
proposed. We will focus on one of them, i.e., thermoelectricity, the direct
conversion of a temperature gradient into an electrical current.
\cite{DiSalvo1999, Vining2008, Dresselhaus2007,Pollock1985,Goldsmid2010}

Nanoscale devices have the power to revolutionise the field of
electronics.\cite{Klein1997, Kubatkin2003, Cuniberti2005, DiVentra2004} It is
clear that their comprehensive and exhaustive experimental exploration not
driven by reliable theoretical predictions will be the result of serendipity
not of scientific research. In the last few decades, Numerical and analytical
tools have been developed to help understanding and designing efficient
molecular devices. Remarkable successes have been scored, but the same tools
for nearly all the most promising molecules, e.g., oxides for photovoltaic
devices, have given mixed results. The general consensus goes in the direction
of investigating the same devices with dynamical methods.\cite{Marques2006}
These allow a careful description of the dynamics of both the electronic
and atomic parts of the device thereby understanding their interplay in
establishing the electrical properties of the molecule. One aspect that is
attracting increasing interest is how to make these molecular devices more
stable. We know that thermal excitations either destabilise the chemical bonds
that attach the molecule to the leads or effectively destroy the lead by
breaking the metal-metal (usually Au) chemical bonds of its
components.\cite{Chen2003} The investigation on how to efficiently remove the
heat locally stored in the device is thus of primary importance towards a
nanoscale electronics.\cite{DAgosta2006c} Here again, time-dependent methods
are the tools of choice, since they allow a fundamental description of the
approach to a steady state \cite{Bushong2005, Sai2007,
Stefanucci2004,Stefanucci2007} (when a steady state is established) than the
standard linear response theories. Finally, nanoscale devices seem to overcome
some of the thermoelectric efficiency problems highlighted in bulk materials. A
careful choice of the materials and an accurate tailoring of their properties
may pave the way to efficient thermoelectric nanodevices.

The efficiency of the thermoelectric energy conversion is related to the so
called figure of merit. This number is defined in terms of the microscopical
quantities of the devices, i.e.,
\begin{equation}
	ZT=\frac{S^2\sigma}{\kappa}T
	\label{figure-of-merit}
\end{equation}
where $S$ is the Seebeck coefficient, $\sigma$ the electrical conductance,
$\kappa$ the thermal conductance, which includes the electronic and crystal
contributions, and $T$ the operating temperature. The figure of merit
determines how efficient the Peltier or Seebeck devices will be:
\begin{equation}
	\eta=\eta_0\frac{\sqrt{1+ZT}-1}{\sqrt{ZT+1}+T_</T_>},
\end{equation}
where $\eta$ is the efficiency and $\eta_0$ is the efficiency of the Carnot
ideal cycle, $\eta_0=(T_>-T_<)/T_>$. Here, $T_>$ is the temperature of the hot
junction, while $T_<$ is the temperature of the cool junction. Typical values
for $ZT$ in actual devices are around 1, and viable devices for technological
applications need to have $ZT$ larger than 2-3.\cite{Vining2009}

Nanomaterials have shown a plethora of interesting phenomena, many of which in
the realm of electric and thermal transport \cite{Cuniberti2005, DiVentra2004,
DiVentra2008}. For application to the thermoelectric technology, certain
nanomaterials like metal nanowires present the interesting feature that
electrical and thermal transport are almost independent, in particular the
electrons are mostly responsible for the electrical transport while lattice
vibrations are mostly responsible for thermal transport. This allows us to
tailor nanoscale devices in which the thermal conductance is minimised while
keeping the electrical conductance constant. On the other hand, quantum
mechanical effects can strongly modify the transport properties of
nanomaterials. As discussed by Hicks and Dresselhaus,\cite{Hicks1993} a key
characteristic for determining the efficiency of the thermoelectric energy
conversion is the electronic density of states of the device. In small
strongly confined systems, a tailored density of states might induce a manyfold
increase of the figure of merit. For this reason there is widespread belief
that nanoscopic devices can show a large thermoelectric figure of
merit.\cite{Hicks1993} There are other reasons for searching for efficient
thermoelectric materials in sub-micron world. We have seen that for large
energy production thermoelectric devices might prove not efficient enough.
However, their efficiency remains almost constant independent of the size of
the device itself, while the efficiency of other energy conversion processes
scales dramatically with size. For this reason, in applications in which
controlling the temperature of a small portion of a larger device (like for
example in cooling down a CPU) is fundamental, thermoelectric devices offer an
economical and efficient solution. In the quest for efficiency, the
properties of a wide range of materials have been investigated. Many have shown
good thermoelectric efficiencies and this implies that in choosing the material
for a certain application other parameters can be adjusted, for example, it
is easily foreseen that silicon nanowires might found wider implementation in
current electronic devices based on silicon. Opaque or transparent materials
might find application in the field of photovoltaics, to increase the total
current output of the device by using the part of the solar energy that is lost
to waste heat.

Many strategies have been proposed to increase the figure of merit of a
thermoelectric device: we can either increase $S$, $\sigma$, or the power factor $S^2\sigma$ or decrease the
thermal conductance $\kappa$. It should become
apparent that the maximisation of the figure of merit in bulk materials is
difficult. Indeed, it is well known that the electrical and thermal
conductances in bulk materials are not independent, but, for example in metals,
they are connected by the Wiedemann-Franz empirical law
\cite{Ashcroft1976}.\footnote{Notice that the Wiedemann-Franz law could be
formulated in terms of conductivities or conductances. For our scope, this
distinction is irrelevant, since the same dimensional factors will multiply the
numerator and denominator of \ref{figure-of-merit}.} This law is based on the
observation that in a metal, electrical and thermal currents are both carried
by electrons. However, little is known about how to calculate the
Seebeck coefficient. The definition of $S$ is quite generic,
\begin{equation}
	S=-\left.\frac{\Delta V}{\Delta T}\right|_{I=0}
	\label{seebeck}
\end{equation}
where $\Delta V$ is the voltage drop at the junctions if a temperature gradient
$\Delta T$ is maintained, in an open circuit configuration, i.e., no electrical
current is flowing. It has been suggested that a direct measure of this
coefficient gives direct evidence of the sign of the carriers, a problem in
complex nanoscale systems.\cite{Reddy2007} This statement is generally true for
bulk materials, but it has been pointed out that the theoretical approach used
to understand the experimental results is based on the assumption that, at
equilibrium, both the thermal and the electrical currents
vanish.\cite{Dubi2009a} This assumption is not valid when we consider nanoscale
systems, i.e., the ones for which we hope to get large figures of merit. It has
been shown that removing this condition allows different signs of the
Seebeck coefficient given the kind of carriers in the device while varying the
working conditions.\cite{Dubi2009a}

In this Perspective, I will discuss a nowadays standard approach to the calculation of the figure of merit, namely the combination of the Landauer's approach to quantum transport and the Onsager's theory of non-equilibrium response. I will propose alternative dynamical methods that can go beyond certain limitation of that standard approach.
The paper is organised as follows. In the following section, we will discuss the Landauer's theory of quantum transport. We will show how one can arrive to simple expressions for the physical parameters entering the figure of merit $ZT$. In section \ref{TDDFT}, we will discuss an alternative method for the calculation of the electrical conductance. In section \ref{STDDFT}, we will turn our attention to the Seebeck coefficient. Finally, in section \ref{thermal_conductance} we will discuss the problem of evaluating the thermal conductance for electrons from a dynamical theory. In this latter case, we cannot provide an answer to this question, but we will discuss briefly two research lines which could lead to a comprehensive dynamical theory of the thermoelectric energy conversion. Throughout the paper we will set $\hbar=k_B=1$, where $k_B$ is the Boltzmann constant.

\section{Landauer's approach to quantum transport}\label{landauer}
A widely used and successful approach to the calculation of the conductance in
nano- and mesoscopic systems is the Landauer's formula.\cite{DiVentra2008} In
this elegant and economic formalism one assumes that the details of the system
are irrelevant far from the nanostructure in some region of space called
“reservoirs”: the only property requested from these reservoirs is that their
cross section be infinite, meaning that an electron entering that region cannot
escape. A fixed bias keeps the two reservoirs polarized. The nanostructure is
connected to the reservoirs through two regions of space which smoothly enter it. The transport properties are completely determined by the
probability amplitude that an electron comes out of a reservoir, crosses the
lead-nanostructure-lead device and is absorbed into the other reservoir. It is
clear that this model contains many assumptions. We should be able to identify
the different regions of interest (reservoir-lead-nanostructure) where
different physics is at play, a separation lacking a solid justification
from a microscopical point of view.\cite{Cini1980} Another basic assumption is
that the many-particle states in the reservoirs are fully determined and do not
mix; we are effectively neglecting particle-particle interaction. Finally the
Landauer's approach cannot explain the nature of out-of-equilibrium states that
are established in the device. In particular, it cannot answer the basic
question, will such a system ever reach a steady state? Other approaches suffer
from similar drawbacks. For example the non-equilibrium Green's function
formalism (also known as the Keldysh formalism),\cite{Kadanoff} although a
powerful technique, assumes the existence of a steady state imposed by
scattering boundary conditions far from the nano-system. The existence of a
steady state seems a reasonable assumption for mesoscopic systems but needs to
be proven for nanoscopic devices or closed system.\cite{Stefanucci2007} Other
ad-hoc models have been proposed in the past but they lack a clear physical
interpretation.\cite{DiVentra2008}

Assuming the Landauer's approach can be applied, we start from the linear out-of-equilibrium Onsager's relations for the currents flowing along the device,\footnote{Here the particle and heat currents are assumed to be perpendicular to the cross section of the device at each position. We expect this to be correct for long nano-wires.}
\begin{eqnarray}
	j&=&L_{00}\Delta \mu+L_{01}\Delta T \nonumber\\
	j_h&=&L_{10}\Delta \mu+L_{11}\Delta T.
	\label{onsager}
\end{eqnarray}
Symmetry considerations lead to $L_{01}=L_{10}$. Landauer's approach to thermal
transport allows to express the Onsager's coefficients $L$'s in terms of the
transmission probability. In linear response with respect to the thermal
gradient $\Delta T$ and the polarisation potential $\Delta \mu$, one obtains
the general expressions
\begin{equation}
	L_{ij}(\mu,T)=\int_{-\infty}^\infty dE (E-\mu)^{i+j} \frac{\partial f(E,T,\mu)}{\partial E}\mathcal{T}(E).
	\label{integrals}
\end{equation}
In (\ref{integrals}), the chemical potential $\mu=(\mu_R+\mu_L)/2$, where
$\mu_R$ and $\mu_L$ are the chemical potentials of the reservoirs,
respectively. By assumption, $\mu_L-\mu_R=\Delta \mu\ll \mu_L,~\mu_R$. The same
is true for the temperature $T=(T_R+T_L)/2$, with $\Delta T=T_L-T_R\ll
T_R,~T_L$. $f(E,T,\mu)$ is the equilibrium Fermi distribution. We assume indeed
that the reservoirs are large and at equilibrium at given temperature and
chemical potential. Moreover it is assumed that the electrons in the reservoir
do not interact. This last assumption is justified by the fact that in metal
the screening length is small. Finally, $\mathcal{T}(E)$ is the transmission
probability for one electron with given energy $E$ to travel across the device.
The theory needs to calculate this quantity to make the set of equations
(\ref{onsager}) and (\ref{integrals}) of any use. We would like to point out
here that while Landauer's approach to transport is valid beyond linear
response, the Onsager's relations are valid only in linear response.

Given the Onsager's relations, and remembering the definition of the different
physical quantities, we obtain the following relations between the microscopic
parameters that enter the figure-of-merit and the integral $L$'s,
\begin{equation}
	\begin{split}
		\sigma&=e^2L_{00}(\mu,T),\\
		S&=\frac{L_{10}(\mu,T)}{eTL_{00}(\mu,T)},\\
		\kappa_e&=\frac{1}{T}\left(L_{11}(\mu,T)-\frac{L_{10}^2(\mu,T)}{L_{00}(\mu,T)}\right).
		\label{coefficients}
	\end{split}
\end{equation}
In (\ref{coefficients}), $\kappa_e$ is the \emph{electronic} contribution to
the thermal conductance, defined by
\begin{equation}
	\kappa_e T= \left.\frac{j_h}{\Delta T}\right|_{j=0}.
\end{equation}
Here, we assume that we can write the total thermal
conductance as $\kappa=\kappa_e+\kappa_l$, where $\kappa_l$ is the crystal
(either phononic or vibronic) contribution to the thermal conductance. This is
valid if the electron-phonon interaction is small. Only then we can
separate these two contributions.

A step further into a complete theory of the thermoelectric energy conversion
is the expression of the thermal conductance $\kappa_l$ in terms of the
transmission probability of a phonon across the device,
\begin{equation}
	\kappa_l=\int_0^\infty d\omega\omega^2 \frac{n(\omega,T)}{\partial T}\tau(\omega),
	\label{phonon_kappa}
\end{equation} 
where $n(\omega,T)$ is the Bose distribution of non-interacting phonon at given
temperature $T$, and $\tau(\omega)$ is the probability that one phonon with
energy $\omega$ could cross the
device.\cite{Blencowe1999,Blencowe2004,Yamamoto2006}

It is clear from (\ref{integrals}), (\ref{coefficients}), and
(\ref{phonon_kappa}) that the only ingredients that we are still missing are
the transmission probabilities $\mathcal{T}(E)$ and $\tau(\omega)$. A lot of
progress has been made to connect these probabilities with the non-equilibrium
Green's function theory and ultimately with electronic structure
calculations.\cite{Meir1992,Jauho1994,Yamamoto2006,Mingo2006} Remarkably, the
transmission probability $\mathcal{T}$ can be written as
\begin{equation}
	\mathcal{T}(E)=\mathrm{Tr}\left[G^r(E)\Gamma_L G^a(E)\Gamma_R \right],
	\label{transmission}
\end{equation}  
where $G^r$ is the retarded Green's function, $G^a(E)=G^r(E)^\dagger$, and
$\Gamma_{L(R)}=i\left(\Sigma_{L(R)}^r-\Sigma_{L(R)}^a\right)$ describe the
coupling, via the self-energies $\Sigma^r$ and $\Sigma^a$ of the left and right
reservoir with the device, respectively. Finally, the non-equilibrium Green's
function is obtained from the Hamiltonian via,
\begin{equation}
	G^r(E)=\left[(E+i0^+)+H_C-\Sigma^r_L-\Sigma^r_R\right]^{-1},
	\label{green-function}
\end{equation} 
where $H_C$ is the Hamiltonian of the central system. The Onsager's
out-equilibrium response theory is exact in the linear approximation. Therefore,
the quality of the results is strictly related to the approximations used to
calculate the Hamiltonian $H_C$ and its eigenvalues. A certain degree of
success has been obtained by using for the central region the Kohn-Sham (KS)
Hamiltonian. \cite{Kohn1965,Giulianivignale,DiVentra2008} To some extent, this
degree of success is unexpected. The KS theory indeed, should not give
access to the spectral properties of the system under investigation. Meanwhile,
those properties are indeed strongly used in the evaluation of the
non-equilibrium Green's functions. Finally, expressions similar to
(\ref{transmission}) and (\ref{green-function}) can be written for the
phonon-transmission probability, with the substitution of $E$ with $\omega^2$,
$H_C$ with $K$ the spring constant matrix, $G^r$ with $D^r$ the phonon
non-equilibrium Green's function.

The fact that one can tackle on the same ground both the vibrational and
electronic contribution to the figure-of-merit is one of the greatest
advantages of this theory. Moreover, its simplicity in expressing all the
quantities of interest in terms of the transmission probability and this one in
term of a microscopic model of the device, is invaluable in making the theory
\emph{predictive}.\cite{Quek2011}

At the same time, to gain a deeper understanding on the physics of
thermoelectric energy conversion, we should be able to build other models and
theories that can extend or confirm the validity of the Landauer/Onsager's
approach to the calculation of the figure-of-merit. Over the years, many novel
approaches have been proposed which are trying to go beyond the actual
state-of-the-art in molecular transport, by, e.g., implementing a {\it GW}
approximation for the calculation of the non-equilibrium Green's function and
apply it to transport in molecular devices\cite{Thygesen2008,Thygesen2008a} or
by building a suitable base for calculating the time evolution of the
electrons.\cite{Bokes2007} In the following we will discuss other ways,
strongly based on Time Dependent Density Functional Theory (TDDFT),\cite{Marques2006} to calculate the Seebeck coefficient
and the electrical conductance by using dynamical theories. Here, one
studies the dynamics of the electrons as they respond to the external
perturbation that generates the electrical and heat flows. In general, these
theories have the possibility to go beyond the linear response, and assess
non-linear effects. The drawback is that the computation of the electron
dynamics is more expensive than any static calculation. At the same time, the
dynamical theories are still in their infancy. This means that we do not have
reliable approximations to the many-body Hamiltonian that allow a correct
evaluation of the dynamics of the observable, although some remarkable progress
has been made through, for example, exact solutions\cite{Maciejko2006} or in
correlating the dynamics of the electrons with the
ions.\cite{Sergueev2005,Lu2012} In this respect, further investigation is
necessary. While these alternative models do allow the evaluation of the
Seebeck coefficient and the electrical conductance, little can be said about
the thermal conductances, especially for the electrons. In Section
\ref{thermal_conductance}, we will shortly address this question and outline
two possible research paths for modelling the thermal electronic conductance
with dynamical theories.

\section{Time-dependent (current-)density functional theory}\label{TDDFT}

Density Functional Theory (DFT) has become a ubiquitous tool in the
investigation of the properties of matter.\cite{Kohn1965,Giulianivignale} The
reasons for this success are manifold and introducing the reader to the field
is well outside the scope of this article. It will suffice to say that DFT
gives in principle access to an exact solution of the
many-body problem, whenever we can focus only on certain quantities, like,
e.g., the single particle density. In its original formulation, DFT deals with
the problem of calculating the ground state energy and single particle density
of a many-body electronic system. Over time, many extensions of the original
idea have been presented: the most interesting for our discussion are the
TDDFT and the Time-Dependent
Current-Density Functional Theory
(TDCDFT).\cite{Runge1984,Ghosh1988,Giulianivignale,Marques2006} As their names
suggest, the two theories deal with the problem of calculating the dynamics of
a many-body system. The result is obtained by mapping the dynamics of the real
system onto the dynamics of a fictitious many-body system of non-interacting
particles. In the case of TDDFT, we are interested in the single-particle
density, $n(r,t)$. Interestingly, TDDFT does not give access to the current
density $j(r,t)$\cite{DAgosta2005a} but it can reproduce the exact total
current.\cite{DiVentra2004a} On the other hand, TDCDFT gives access to the
exact single-particle current density, and via the continuity equation, to the
single-particle density. In this respect TDCDFT is a more general theory than
TDDFT. We would like to point out that DFT and TDDFT should be able to deal
with the dynamics of strongly correlated systems. In this direction, remarkable
results have been obtained in the solution of the Kondo problem, see for
example \cite{Stefanucci2011} and references therein. One of the advantages of
this theory over the standard non-equilibrium Green's function formalism is
that it avoids the separation of the system intro three (or more) parts. In
general, indeed for the non-equilibrium Green's function we need to identify
the left and right reservoirs and connect them with the central region. In most
of the cases, this procedure lacks a clear physical
interpretation\cite{Cini1980} and may produce spurious effects. On the other
hand, in TDCDFT one can investigate the full dynamics of the system without
reverting necessarily to this separation.
\cite{Stefanucci2008,Bushong2005,Kurth2005}

It should also be clear that TDCDFT is built to treat problems where electrical
transport is important. By reproducing the single-particle current for given
external potential, the electrical conductance, $\hat \sigma$ can be obtained
\emph{directly} from its microscopical definition
\begin{equation}
	\vec j(r,t)=\int dr' \hat \sigma(r,r')\cdot \vec E(r',t).
	\label{conductance}
\end{equation}
Since the electrical field $\vec E(r,t)$, is an external input,
(\ref{conductance}) allows the complete determination of the conductance
tensor, if the current density can be calculated. Moreover, the theory does not
suffer from any problem of dividing the device/reservoir region. In principle,
we can treat the motion of the electrons exactly in the presence of an external
polarizing potential.\cite{Stefanucci2004}

\section{Calculation of the Seebeck coefficient}\label{STDDFT} 
We will now turn
our attention to the calculation of the Seebeck coefficient. It should be
apparent from its definition (\ref{seebeck}) that the calculation and
measurement of the Seebeck coefficient are a difficult task. Indeed, in the experiments we combine
the ability to control the temperature gradient with the measurement of small
bias drops while maintaining an open circuit configuration. These technical difficulties
reflect into a large uncertainty in the values of the Seebeck coefficient which
is usually the result of indirect measurements which in turn require a good
level of theoretical understanding. From a theoretical point of view is then of
the uttermost interest to have solid and reliable ways to calculate the Seebeck
coefficient for many different materials. In this respect, the theory we have
exposed in Section \ref{landauer} has proven an invaluable tool. Nonetheless,
we must find an alternative formulation that will teach us the strengths and
weaknesses of the Landauer's approach and provide novel tools to pursue our
quest of efficient thermoelectric materials. Ideally, we would like to revert
to the basic definition of the Seebeck coefficient, $S=-\Delta V/\Delta T$ at
vanishing electrical current. $\Delta V$ is the response of the system to the
external thermal gradient, $\Delta T$. We then need a theory able to describe
the dynamics and steady state of a quantum mechanical system coupled to two
external baths, or, more in general, in contact with an external environment.

The theory of open quantum systems has been devised with this aim in
mind.\cite{Feynman1963,Gisin1981,Caldeira1983,Diosi1997,Strunz1996,Gardiner2000,
Weiss2007} The general starting point is the Hamiltonian of the composite
system: device plus external environment. Generally we can write that
Hamiltonian as
\begin{equation}
	H=H_C+H_E+H_{coup}
\end{equation}
where $H_C$ is the Hamiltonian of the isolated system, $H_E$ the Hamiltonian of
the environment, and $H_{coup}$ describes the coupling between the two.
The aim of the theory is to ``fold'' the degrees of freedom of the environment
and obtain an effective equation of motion for the system dynamics. If we
assume the coupling has the form
\begin{equation}
	H_{coup}=\lambda V\otimes B,
\end{equation}
where $V$ and $B$ are system and environment operators, respectively, and
$\lambda$ is a small coupling parameter, a standard procedure leads to the
equation of motion for the ``state"\footnote{It is important to point out that
the state $\Psi$ generally does not describe a true quantum
trajectory.\cite{Biele2012}} $|\Psi\rangle$ of the
system\cite{Biele2012,Gaspard1999,Gardiner2000,vanKampen}
\begin{equation}
	\begin{split}
	i\frac{\partial}{\partial t}|\Psi(t)\rangle=&-iH_C|\Psi(t)\rangle\\
	 &+ \lambda^2\int_0^t d\tau C(\tau)V^\dagger(\tau)e^{-iH_C\tau}V|\Psi(\tau-t)\rangle\\
	&+\lambda\eta(t)V|\Psi(t)\rangle,
\end{split}
\label{SSE}
\end{equation}
where $C(t)=\mathrm{Tr}_{E}\left[\rho_E^{eq}B(t)B^\dagger(0)\right]$ is the
environment correlation function. We assume that the environment is maintained in the thermal equilibrium
described by the statistical operator
\begin{equation}
	\rho^{eq}_E=\frac{e^{-\beta H_E}}{\mathrm{Tr}_E e^{-\beta H_E}},
\end{equation}
and $\eta(t)$ is a complex coloured noise with the statistical properties\cite{Gardiner2000} 
\begin{equation}
	\overline{\eta(t)}=0,~\overline{\eta(t)\eta(0)}=0,~\overline{\eta(t)\eta^*(0)}=C(t).
\end{equation}
Here with the $\overline{\cdots}$ we indicate the average taken after many
realisations of the noise.

The stochastic Schr\"odinger equation (SSE) (\ref{SSE}) is able to reproduce
the dynamics of the density matrix for the
system,\cite{Gaspard1999,Strunz2000,DeVega2005,Biele2012}
\begin{equation}
	\rho=\frac{\overline{|\Psi\rangle\langle\Psi|}}{\overline{\langle\Psi|\Psi\rangle}}.
\end{equation} 
Interestingly, it is the average over the many realisations of the noise that
makes this density matrix not pure.

A standard approximation consists in replacing the correlation function $C(t)$
with a Dirac $\delta(t)$ function. This corresponds to neglecting the time
correlation of the bath, effectively assuming the environment has no
memory - a Markov approximation. Within the Markov approximation, the SSE
assumes the simplified form
\begin{equation}
	\begin{split}
	i\frac{\partial}{\partial t}|\Psi(t)\rangle=&-iH_C|\Psi(t)\rangle\\
	 &+ \lambda^2V^\dagger V|\Psi(t)\rangle+\lambda\eta(t)V|\Psi(t)\rangle.
	\label{markov_sse}
	\end{split}
\end{equation}  
Using the definition, we can derive a simple form for the dynamics of the
master equation,
\begin{equation}
	\frac{d\rho}{dt}=-i\left[H,\rho\right]-V\rho V^\dagger+\frac12 V^\dagger V\rho+\frac12 \rho V^\dagger V.
	\label{lindblad}
\end{equation}
It can be proven that this is the most general memoryless master equation valid
up to second order in $\lambda$ that preserves the positivity of the density
matrix.\cite{Lindblad1976}

The equations (\ref{markov_sse}) and (\ref{lindblad}) can be easily generalised
to the case of the system being in contact with more than one bath operator
$B$. In doing so, we can assume that each of these baths corresponds to a
different thermal equilibrium (for example with different temperatures) and
then study the dynamics of the system under this now out-of-equilibrium conditions.
This idea has been explored partially to investigate the thermoelectric energy
conversion with (\ref{lindblad}).\cite{Dubi2009a} It is important to point out
that in principle, in the Markov approximation, the detailed balance equation
that will ensure the establishment of thermal equilibrium when the temperatures
of the baths are the same, is not fulfilled.\cite{Biele2012} Moreover, it is
not clear how to treat particle-particle interaction in this formalism, since
that in general will make the Hamiltonian state-dependent. The derivation of
the master equation in this case appears more subtle and care is needed to
avoid spurious effects.\cite{DAgosta2008a} For this reason, a TDCDFT scheme for
this open quantum system has been recently
developed that can go beyond the limits of the
density matrix formalism.\cite{DiVentra2007,DAgosta2008a} 
This scheme, by coupling the strength of a DFT approach
with the theory of open quantum systems, could help in establishing a framework
for the investigation of the Seebeck coefficient of materials. Indeed, being
able to couple the device of interest with two baths at different temperatures,
we can study the formation of the potential difference $\Delta V$ in time, by
using the Poisson equation, since the electrons in the system will try to
oppose to the thermal gradient by piling up closer to the coolest bath. At
equilibrium, no electronic macroscopic current will be present in the device,
and we can use the definition (\ref{seebeck}) to evaluate the Seebeck
coefficient.\cite{Dubi2009a} The same theory can be used to investigate the
validity of the Fourier law at the nanoscale.\cite{Dubi2009}

\section{Thermal conductance of electrons}\label{thermal_conductance}

As we have seen in the evaluation of the figure of merit, the Seebeck
coefficient and the electrical conductance can be calculated by using dynamical
methods thus offering an alternative description to the Landauer formalism.

The situation is less clear with respect to the calculation of the thermal conductance. In this
case, there is not a dynamical formulation of the problem in terms of a density
functional approach. The reason for this situation is simple: The energy
current $j_h$ cannot be written in terms of the electrical (or density)
current. Indeed, we can define the energy current across a given section of the
system as
\begin{equation} 
	j_h=-\frac{d}{dt}\langle H_S\rangle 
\end{equation} 
where $H_S$ is the energy stored in the volume enclosed by the surface we are
considering. For example, if we can separate the system into left, right and
central region, where the left and right act as energy ``reservoirs'', $H_S$ is
the energy stored in one of these two reservoirs. We could then write a similar
equation for the energy coming from the other reservoir. Combining the two, we
can calculate the energy flow in the system. A naive approach would simply
suggest to calculate the energy starting from the DFT wavefunctions, in a
fashion similar to what has been done in the past to evaluate the
non-equilibrium Green's function. However, this operation contains a certain
number of approximations, many of which are not under control. Indeed, the
theorem of DFT (either the static or dynamical) cannot give access to the
energy of the excited states. For this reason, the concept of the total energy
stored in a part of the system will bear little meaning in a DFT formulation of
the problem.

We foresee two ways out of this impasse. On the one hand, one can think of
building an ad-hoc density functional theory that provides the exact energy or
thermal current. To do so, we need to identify a suitable potential
that is connected to the energy current, responsible to generate that current
in the many-body system. If that potential is found, its existence is doubtful
to say the least, then we can hopefully prove that the mapping between the
potential and the energy current is one-to-one. If this can be done, and a
suitable KS scheme be built, we will have an exact formulation of the
problem in terms of non-interacting particles and we can compute the energy or
the thermal current from first principles.

However, one can build some approximate model to the electron thermal
transport. In this respect a hydrodynamical formulation of the electronic
transport, would provide a set of equations which includes the thermal
conductance and the specific heat of the electron liquid. See for example
Ref. \cite{DAgosta2006c} for the formulation of the theory for a quantum point
contact. In Ref. \cite{DAgosta2006c} the hydrodynamical equations were used to
calculate the electronic temperature in the nanojunction. However, if we do
assume that temperature can be either modelled or measured independently, these
equations will be sufficient to calculate the thermal conductance of the
electrons. An \emph{ab-initio} theory here will enter in defining the viscous
coefficient of the electron liquid.

In a hydrodynamic approach to quantum transport, we define the {\it stress
tensor} starting from the velocity of the fluid, $\vec v(r,t)=\vec
j(r,t)/n(r,t)$,
\begin{equation}
\pi_{i,j}=\eta
\left(\nabla_{i}v_{j}+\nabla_{j}v_{i}-\frac{2}{3}\delta_{i,j}\nabla_{k}v_{k}\right)
\label{pixcstatic}
\end{equation}
where $\eta$ is a real coefficient (the viscosity) that is a functional of the
density.\cite{Tokatly1999,DAgosta2006c,DAgosta2006d} We point out that
(\ref{pixcstatic}) is in fact a particular case of a general stress tensor with
memory effects taken into account.\cite{Conti1999a,Vignale1996,Tokatly2005b}
The parameters of the liquid can be build from a microscopic theory of the
low-energy excitations of the system under investigation. Unfortunately, it
turns out that their detailed dependence on the geometry and dimensionality of
the system is complex.\cite{DAgosta2006d} It is clear that further
investigation in this direction is needed. In terms of this stress tensor, the
continuity equation and the equation of motion for the velocity field are
written as,
\begin{equation}
	\begin{split}
D_{t}n(r,t)=&-n(r,t)\nabla \cdot \vec v(r,t),\\
mn(r,t)D_{t}v_{i}(r,t)=&-\nabla_{i}P(r,t)+\nabla_{j}\pi_{i,j}(r,t)\\
&-n(r,t)\nabla_{i}V_{ext}(r,t).
\label{completeNS}
\end{split}
\end{equation}
where $D_t=\partial/\partial_t+v(r,t)\cdot\nabla$ is the convective derivative,
$m$ the electron mass, $P(r,t)$ is the pressure, and $V_{ext}(r,t)$ is a
time-dependent external potential. The second equation is nothing else than the
Navier-Stokes equation for the a \emph{viscous}
fluid.\cite{Landau6,DAgosta2006c} It is interesting to point out that for a
nanoscale system, for electronic low-energy excitations, this theory is
equivalent to the many-body formulation.

To study the thermal transport, we need to complement (\ref{completeNS}) with an
equation for the energy flow. For a quantum point contact, a nanoscale
constriction between two large metallic reservoirs, one
obtains,\cite{DAgosta2006c}
\begin{equation}
\pi_{i,j}(r)\partial_j
v_i(r)+\nabla\cdot [k(r) \nabla T_e(r)]=c_{V}(T_{e})\vec v(r)\cdot \nabla T_e(r),
\label{heatequilibrium}
\end{equation}
where $T_e$ is the electronic temperature, $k(r)$ is the diffusion constant and
$c_V$ is the specific heat at fixed volume of the electron gas. Obviously, in
writing Eq.~(\ref{heatequilibrium}) we have assumed that some thermodynamic
quantities like temperature and entropy for an electron liquid flowing in a
nanostructure can be defined. This is a much debated point, and obviously we do
not have a general solution for it. However, here we argue that the electron
temperature may be defined as the one ideally measured by a probe weakly
coupled to the system and in local equilibrium with the latter.
\cite{Dubi2009a,DiVentra2008} The set of hydrodynamical equations has been
used to study the thermal heating of a nanoscale device. They do predict novel
and interesting phenomena, like a cooling of the phonon modes due to the
electron heating.\cite{DAgosta2006c} This result has been beautifully confirmed
by experience.\cite{Huang2007}

I am personally amazed that the equation of motion for a many-body system does
reduce to a form similar to the Navier-Stokes equations of a classical liquid.
The quantum mechanical effects are still embodied by the pressure term $P(r)$. For
this reason, the solution of these Navier-Stokes equations is somehow more
difficult than the already demanding solution of the classical equations.
Finally, it is quite surprising that TDCDFT arrived at this same conclusion
well before.\cite{Vignale1996,Conti1999a} At the same time, TDCDFT seems to
suggest natural mechanisms for energy relaxation,\cite{Wijewardane2005} and
recently a thermodynamical interpretation of the KS energy has been
given, as the maximal work that can be extracted from the
system.\cite{DAgosta2006} Starting from these considerations, it does appear
possible that TDCDFT can be used to describe some of the physical process
related to energy dissipation and transfer.

Both formulations of the problem of calculating the electron thermal transport
will require the solution of long-standing problems. For example, in a density
functional theory for the thermal current, we will very soon discover that a
thermal current is generated by almost any external potential, not only by a
temperature imbalance. Therefore, how can a given thermal current be in a
one-to-one mapping with a certain potential? The hydrodynamical formulation can
be equally as challenging. For example, what is the range of validity of the
equations of motion. For sure, at short wavelengths, the electrons do not behave
as a liquid. Moreover, at the nanoscale, a few very fundamental equations of
classical heat transport, like the Fourier equation, are not valid. The
analysis of thermal transport at the nanoscale thus should include a wider
analysis on the validity of assumptions made on the dynamics of the system.

We have left out the problem of phonon/vibron thermal transport, and the
coupling between the electrons and the vibrons. Presently, besides doing a
full dynamics of the coupled electronic and atomic motions, we do not see an
alternative to the static calculations performed starting from
(\ref{phonon_kappa}), and improving the calculation of the Green's function via
perturbation theory. A lot of effort has been put into the calculation of the
phonon thermal conductance, for example in silicon based nano-structures,
\cite{Donadio2009,Donadio2010} via either molecular dynamics or the Boltzmann
transport equation. While these techniques have been proven useful in
calculating the thermal conductance, it also remains difficult to obtain full
convergence of the dynamics especially when the system is maintained out of
equilibrium. On the other hand, it is true that (\ref{phonon_kappa}) is limited
to the harmonic approximation for the phonon spectrum. This approximation
appears to be valid only for low energy vibrations--since the interaction
parameter vanishes with vanishing phonon energy--which are also responsible for
the largest contribution to thermal transport.
  
\section{Conclusions}
The investigation of the thermoelectric properties of materials poses
interesting challenges because it does require the treatment, on equal footing,
of electrical and thermal transport. So far the only theory that has succeeded
in doing so is the linear response theory developed by using the Onsager's
relations coupled with the Landauer's approach to thermal transport. While
successful for many systems, it is necessary to have alternative theories able
to describe the same systems and go beyond the non-equilibrium linear response
regime. In this article, I have discussed the state-of-the-art of one of the
alternatives, namely a dynamical approach to the calculation of the
thermoelectric coefficients. While the theory is based on solid ground for the
Seebeck coefficient and the electrical conductance, the thermal conductance,
both for the electronic and atomic systems, is more difficult to access with
first principle theories. I have sketched a couple of possible research routes
that could give access to this important piece of information.

\section{Acknowledgments} 
The author acknowledges support from MICINN (FIS2010-21282-C02-01 and PIB2010US-00652), Grupos Consolidados UPV/EHU del Gobierno Vasco (IT-319-07), and ACI-Promociona (ACI2009-1036), the financial support of the CONSOLIDER-INGENIO 2010 ``NanoTherm" (CSD2010-00044) of the MICINN and is thankful for its hospitality to the Physics
Department of the King's College London.

\footnotesize{
\bibliography{library} %your .bib file
\bibliographystyle{rsc} %the RSC's .bst file
}

\end{document}